\documentclass[a4paper,11pt]{article}
\usepackage{pos}

\usepackage{subfigure,psfrag}

\title{Center-symmetric Landau gauge, the deconfinement transition and the gluon propagator as seen in lattice QCD}
 \ShortTitle{Center-symmetric Landau gauge on the lattice}

\author[a]{Duifje Maria van Egmond}
\author[b]{Orlando Oliveira}
\author[c]{Urko Reinosa}
\author[d]{Julien Serreau}
\author*[b]{Paulo J. Silva}
\author[e]{Matthieu Tissier}

\affiliation[a]{ICTP South American Institute for Fundamental Research Instituto de Física Teórica,\\
UNESP - Univ. Estadual Paulista Rua Dr. Bento Teobaldo Ferraz 271, 01140-070, São Paulo, SP, Brazil}

\affiliation[b]{CFisUC, Department of Physics, University of Coimbra, P-3004 516 Coimbra, Portugal.}

\affiliation[c]{Centre de Physique Théorique, CNRS, Ecole polytechnique, IP Paris, F-91128 Palaiseau, France}

\affiliation[d]{Université Paris Cité, CNRS, Astroparticule et Cosmologie, F-75013 Paris, France.}

\affiliation[e]{Sorbonne Université, CNRS, Laboratoire de Physique Théorique de la Matière Condensée, 75005 Paris, France.}

\emailAdd{psilva@uc.pt}

\abstract{We address the lattice computation of the gluon propagator in the center-symmetric Landau gauge. After discussing a proper lattice implementation of the center-symmetric Landau gauge, we compare the lattice data with analytical results, and we identify various signatures of center symmetry breaking.}

\FullConference{The XVIth Quark Confinement and the Hadron Spectrum Conference (QCHSC24)\\
 19-24 August, 2024\\
 Cairns Convention Centre, Cairns, Queensland, Australia\\}

\begin{document}
\maketitle

\section{Introduction and motivation}

The understanding of the phase diagram of QCD has been the goal of a rich and intense research activity. Following several heavy ion experimental programs,  recent theoretical studies focus on the properties of QCD at finite temperature and density.

At low temperatures and densities, quarks and gluons are confined within hadrons. On the other hand, quarks and gluons behave as free quasiparticles at sufficiently high temperatures or densities. At zero density, the transition between these two regimes is signaled by the Polyakov loop $L$, which is the order parameter for the confinement-deconfinement transition. Below the critical temperature $T_c$, and thanks to the center symmetry, the Polyakov loop takes null values, whereas above $T_c$ center symmetry is broken and $L>0$. If $T_c \simeq 270$ MeV for pure gauge theories,  the inclusion of dynamical quarks lowers this value to $T_c \sim 170$ MeV.

In QCD, propagators of fundamental fields encode crucial information about non-perturbative
phenomena, such as confinement, deconfinement and chiral symmetry breaking. Recent studies of gluon, ghost, and quark propagators in Landau gauge (see, for example, \cite{silva2014, silva2016, Oliveira2019, Paiva2023,Cucchieri:2010lgu,Fischer:2010fx,Maas:2011ez,Mendes:2015jea,Aouane:2011fv,Aouane:2012bk,Cyrol:2017qkl} and references therein) showed that the form factors associated to the two-point functions of QCD are sensitive to the deconfinement phase transition. Notwithstanding, recent works by some of us \cite{egmond1, egmond2, scipost} using the so-called \textit{center-symmetric Landau gauge} allowed to formally construct local order parameters for this phase transition.

Here we discuss preliminary lattice results for the center-symmetric Landau gauge. We start by defining this gauge in the continuum and discuss its lattice implementation. Then  some preliminary results for the link average and for the gluon propagator are reported.

\section{Center symmetric Landau gauge}

In the continuum, the following condition 
\begin{equation}
  D_{\mu}[\bar{A}](A_{\mu}-\bar{A}_{\mu})=0
\end{equation}
 where the background covariant derivative is
\begin{equation}
 D_{\mu}[\bar{A}] = \partial_{\mu} - ig \left[\bar{A}_{\mu},\ldots \right],
\end{equation}
defines the family of background Landau gauges. The so-called center-symmetric Landau gauge is characterized by
a center symmetric background configuration $\bar{A}_{c,\mu}$, which is given by
\begin{equation}
  \bar{A}_{c,\mu}=\frac{T}{g} \bar{r}_j t_j \delta_{\mu 0}.
\end{equation}
For the  $SU(3)$ case considered in this work we have $j=3$, with $\bar{r}=4\pi/3$ and $t_j=\lambda_j/2$.
Therefore the final expression for the center-symmetric Landau gauge is
\begin{equation}
 D_{\mu}[\bar{A}_c] (A_{\mu})= \partial_{\mu} A_{\mu} - ig \left[\bar{A}_{c,0},A_0\right].
\end{equation}

On the lattice, where the temporal direction is now labelled by the index 4, the center-symmetric Landau gauge can be defined through the maximization over the gauge orbits of the gauge fixing functional
\begin{equation}
  F=\sum_{x,\mu}\mbox{Re Tr} \left[ g_{c}^{\dagger}(\mu)  U_{\mu}(x)  \right]
  \label{f}
\end{equation}
where  $g_{c}^{\dagger}(\mu)=g_{c}^{\dagger}(x+\mu)g_c(x)$. The matrix $g_{c}^{\dagger}(\mu)$ can be written as
\begin{equation}
  g_{c}^{\dagger}(\mu)=e^{i a T \, \bar{r}_j t_j \delta_{\mu 4}}=e^{i a g \bar{A}_{c,\mu}\delta_{\mu 4}}
\end{equation}
where $a$ is the lattice spacing, the temperature $T$ is defined from $aT=1/L_t$,  with $L_t$ being the number of points in the temporal direction. Note that besides the usual (periodic) gauge transformations, we can also consider \textit{center transformations}
\begin{equation}
U_\mu(x)\to  g(x)U_\mu(x)g^\dagger(x+\hat\mu)\equiv U_\mu^g(x)\,,\label{eq:V_transfo}
\end{equation}
that are periodic in the time direction but only modulo an element of the center of SU($3$), for example
 \begin{equation}
g(x+L_4\hat{4})=e^{i2\pi/3}g(x)\,.\label{eq:modulo}
\end{equation}

Center-symmetric Landau gauge fixing can be done in a pretty similar way to the standard Landau gauge. In this work we rely on a Fourier-accelerated steepest descent method \cite{Davies} to handle the optimization problem.

\section{Center invariance}

A remarkable property of center-symmetric Landau gauge functional is that the gauge-fixing functional F  (eq. \ref{f}) is invariant under the particular center transformation
\begin{equation}
  \tilde{g}(x)=e^{i\pi\frac{\lambda_4}{2}}e^{i\pi\frac{\lambda_1}{2}}e^{-i\frac{x_4}{L_4}\pi\left(\lambda_3+\frac{\lambda_8}{\sqrt{3}}\right)}.
  \label{gtilda}
\end{equation}
Note that, although this transformation is periodic modulo a center element
    \begin{displaymath}
    \tilde{g}(x+L_t\hat{4}) = e^{i\frac{2\pi}{3}}\tilde{g}(x) \, ,
    \end{displaymath}
the gauge-fixing functional F is invariant, and $U^g$ still maximizes F. Correspondingly, under this transformation the Polyakov loop also changes according to  $P_L \to   e^{- i\frac{2\pi}{3}} P_L$.

 \section{Features of the center-symmetric Landau gauge}

 Here we report on lattice results obtained with numerical simulations of the Wilson gauge action for $\beta=6.0$ on a $64^3 \times L_t$ lattice volume, for a temperature below $T_c$ ($T=243$MeV, $L_t=8$) and another temperature above $T_c$ ($T=324$MeV, $L_t=6$).

 Left-hand plots of Figures \ref{A38belowTc} and \ref{A38aboveTc} show the lattice data for $ag(A^3_4(x=0),A^8_4(x=0))$, taken from each lattice configuration, in the three different $Z_3$-sectors that can be obtained after applying the transformation (\ref{gtilda}) to the lattice configurations of the ensemble. These transformations correspond to $\pm 2\pi/3$ rotations in the $(A^3,A^8)$ plane. The gluon field has been computed using 
   \begin{equation}
a g A^a_\mu (x + a \hat{e}_\mu / 2) ~ = ~ 2 \mbox{Im}\mbox{Tr} \Big[ t^a U_\mu (x) \Big]+ \mathcal{O}(a^3).
   \label{gluonfield}
\end{equation}
While below $T_c$ the data fluctuates around (0,0), above $T_c$ each of the different data sets seem to divide the plane into 3 regions aligned with the angles 0, 2$\pi$/3 and  -2$\pi$/3. As expected, by rotating back the data sets to the 0 sector, all sets match each other --- see right-hand plots.

\begin{figure}[h] 
\vspace{0.55cm}
   \centering
   \subfigure[Original data.]{ \includegraphics[width=0.42\textwidth]{plots/Amu3_xzero_38_T243MeV.eps}  \label{A38originalbelowTc}} \qquad
   \subfigure[Rotated data.]{   \includegraphics[width=0.42\textwidth]{plots/Amu3_xzero_38_T243MeV_rotates.eps} \label{A38rotatedbelowTc}}
  \caption{Plotting $(A^3_4(0),A^8_4(0))$ --- below $T_c$:  $64^3\times8$, T=243 MeV.}
   \label{A38belowTc}
\end{figure}

 \begin{figure}[h] 
\vspace{0.55cm}
   \centering
   \subfigure[Original data.]{ \includegraphics[width=0.42\textwidth]{plots/Amu3_xzero_38_T324MeV.eps}  \label{A38originalaboveTc}} \qquad
   \subfigure[Rotated data.]{   \includegraphics[width=0.42\textwidth]{plots/Amu3_xzero_38_T324MeV_rotates.eps} \label{A38rotatedaboveTc}}
  \caption{Plotting $(A^3_4(0),A^8_4(0))$ --- above $T_c$:  $64^3\times6$, T=324 MeV.}
   \label{A38aboveTc}
\end{figure}

 In the continuum, there is a prediction for the gluon field such that   $\beta_T \langle  g A^3_4(x) \rangle = \frac{4\pi}{3}$, see \cite{egmond1, egmond2}, which becomes $\langle a g A^3_4(x) \rangle = \frac{4\pi}{3L_t}$.
This can also be studied through the link average:
    \begin{eqnarray}
      \frac{\langle U_4(x)\rangle }{({\rm det}\,\langle U_4(x)\rangle)^{1/3}}=e^{-\frac{i}{L_4}\frac{4\pi}{3} \frac{\lambda^3}{2}}= \left( 
\begin{array}{ccc}
e^{- i\frac{2\pi}{3 \, L_4}} & 0 & 0 \\
0 & e^{i\frac{2\pi}{3 \, L_4}} & 0 \\
0 & 0 & 1
\end{array}
\right)
    \end{eqnarray}

    For our simulation in the symmetric phase, T=243 MeV, the lattice averages for the diagonal elements of the link average and the corresponding statistical errors are given by
 \begin{eqnarray}
    0.96726(15) &-& i\, 0.25422(56)  \\
    0.96715(13) &+& i\, 0.25456(48)  \\
   0.999800(26) &-& i \,0.00036(60) 
 \end{eqnarray}
 whereas  non-diagonal elements are zero within errors. The theoretical prediction for the first element is  $0.9659258263 - i \, 0.2588190451$, so the numerical values are pretty close to it. Charge conjugation imposes that the second diagonal element is the complex conjugate of the first element, while we find the third element to be close to 1. So the numerical simulations comply with symmetry constraints, meaning that the center symmetry is not broken.
    
 For our simulation above $T_c$,  T=324 MeV, the diagonal elements of the link average are
  \begin{eqnarray}
   0.985563(39)  &-& i\, 0.16827(19)\\
   0.985569(41)  &+& i\,  0.16817(18)\\
   1.000362(20)  &+& i \,  0.00010(15)
 \end{eqnarray}
Similarly to the symmetric phase, the non-diagonal elements are zero within errors. However, since the theoretical prediction for first element is now $0.9396926208 - i \,0.3420201433$, clearly the numerical simulations deviate from the symmetry constraints, signalling a breaking of the center symmetry.

\begin{figure}
\begin{center}
  \psfrag{XXXXXXX}{}
  \includegraphics[width=0.78\textwidth]{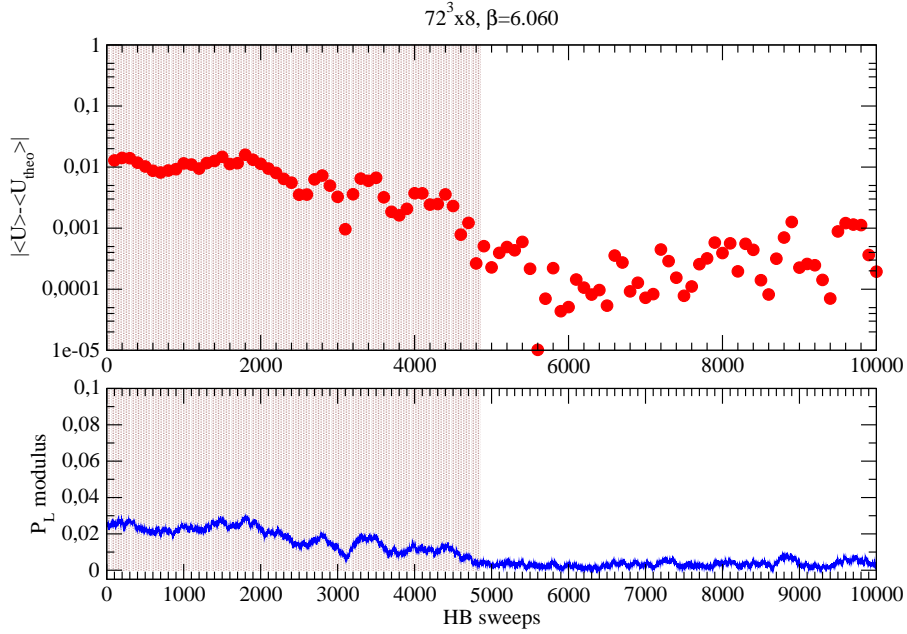}
  \end{center}
\caption{Monte Carlo history of a lattice simulation close to $T_c$.}
\label{near}
\end{figure}

This can also be seen in the Monte Carlo history of a lattice simulation near the critical temperature $T_c$ taken from \cite{silva2016} --- see Fig.\ref{near}. Due to the finite volume, the simulation is able to change from one phase to another. It starts in the deconfined phase, where the modulus of the Polyakov loop $|P_L|$ is non-zero, and goes through the confined phase around sweep number 5000, characterized by a vanishing $|P_L|$. We also plot the modulus of the difference between the numerical value of the link average and the corresponding theoretical prediction: this quantity is finite for the deconfined phase, as expected, and approaches zero for the confined phase.

\section{Gluon propagator}

In the center-symmetric Landau gauge, and for color indices 3 and 8, the tensor structure of the gluon propagator becomes the same as in the standard Landau gauge
\begin{equation}
D^{ab}_{\mu\nu}(\hat{q})=\delta^{ab}\left(P^{T}_{\mu\nu} D_{T}(q_4,\vec{q})+P^{L}_{\mu\nu} D_{L}(q_4,\vec{q}) \right) \nonumber
\end{equation}
which means the extraction of the longitudinal and tranverse components of the gluon propagator can proceed in a very similar way as we do for the standard Landau gauge \cite{silva2014}.

\begin{figure}

  \centering

  \psfrag{XXXXXXX}{}
   \subfigure{}{\includegraphics[angle=-90,width=0.42\textwidth]{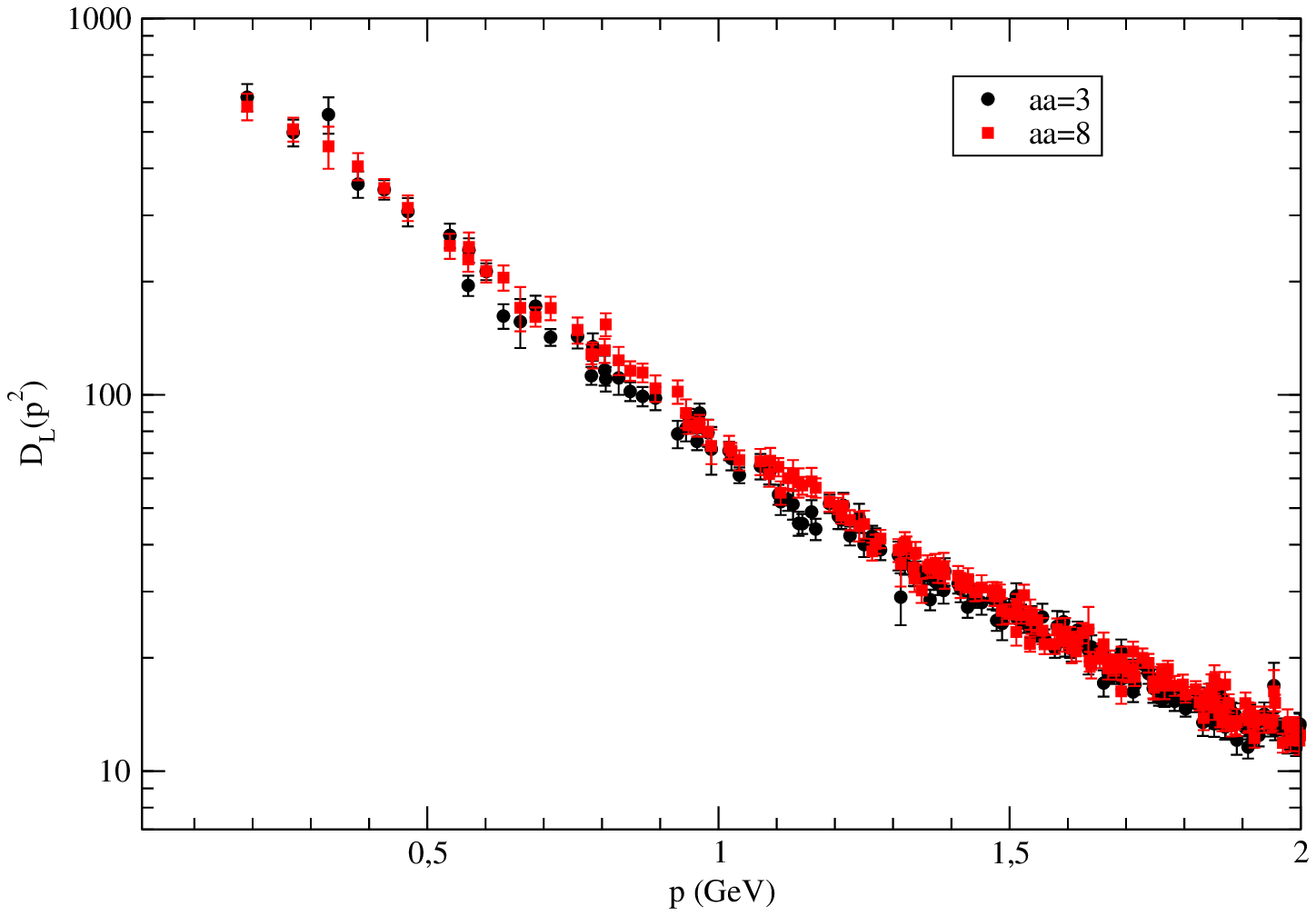}}
   \qquad
   \subfigure{}{\includegraphics[angle=-90,width=0.42\textwidth]{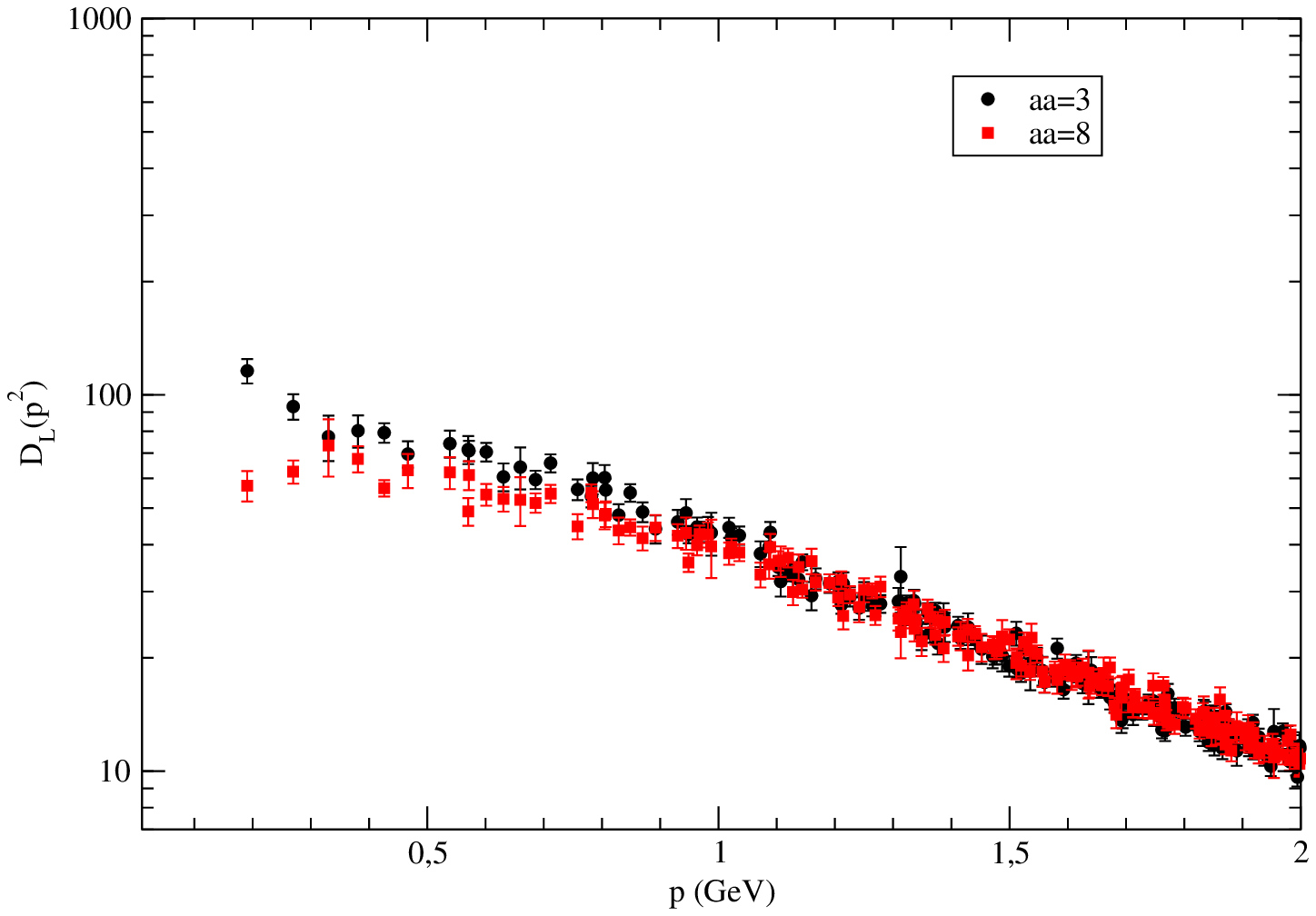}}

    \subfigure{}{\includegraphics[angle=-90,width=0.42\textwidth]{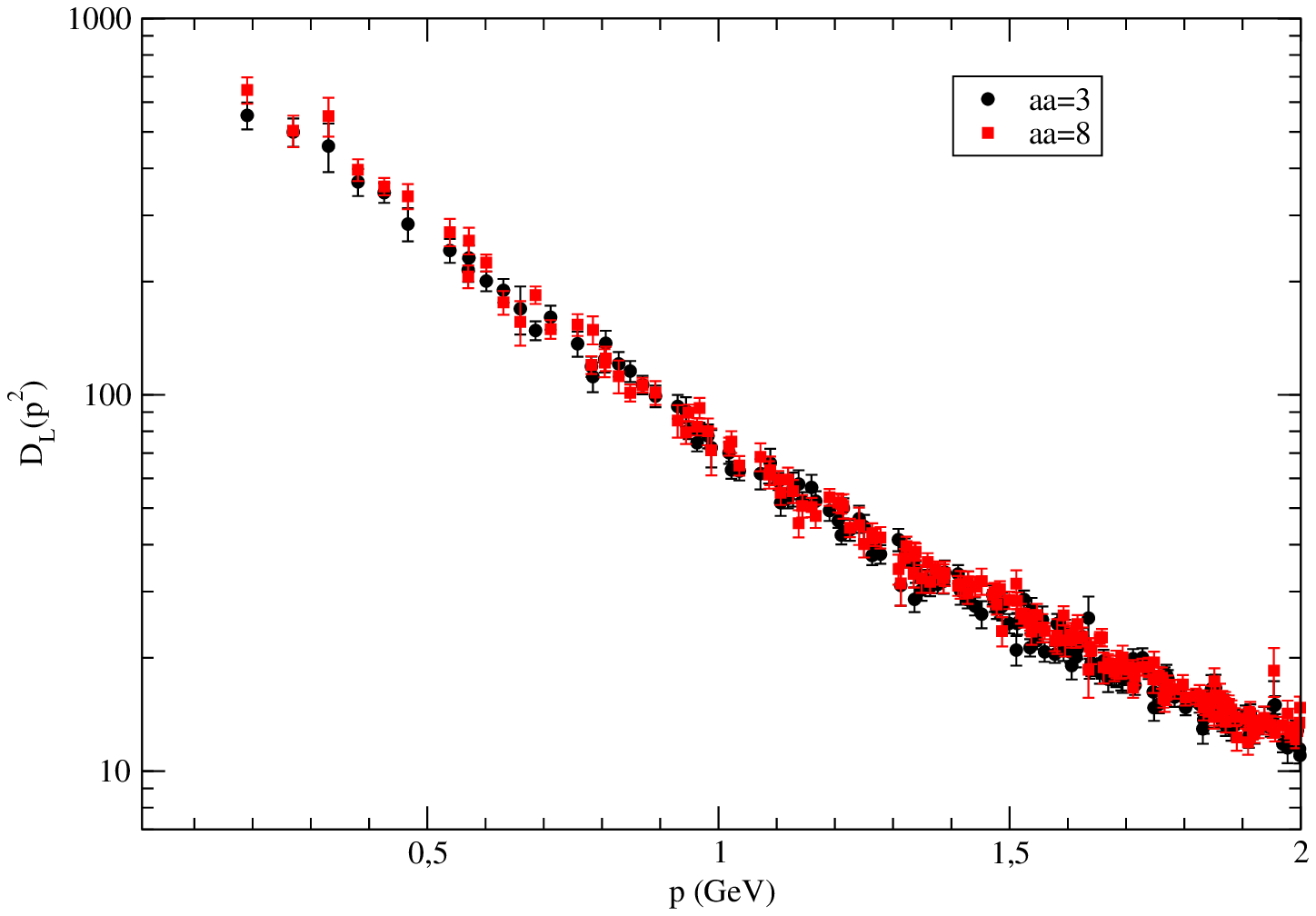} }
 \qquad
 \subfigure{}{\includegraphics[angle=-90,width=0.42\textwidth]{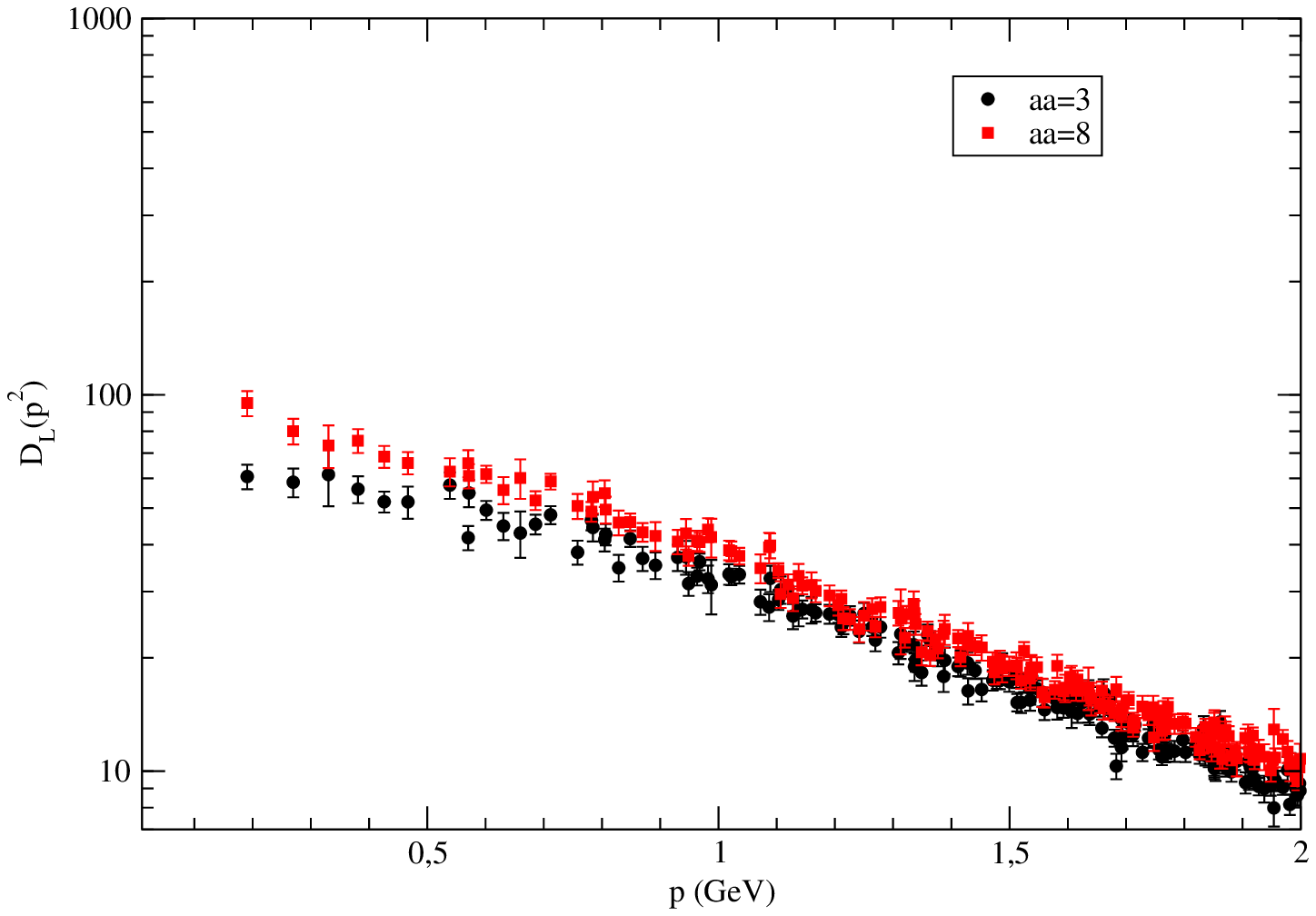}}

\subfigure{}{\includegraphics[angle=-90,width=0.42\textwidth]{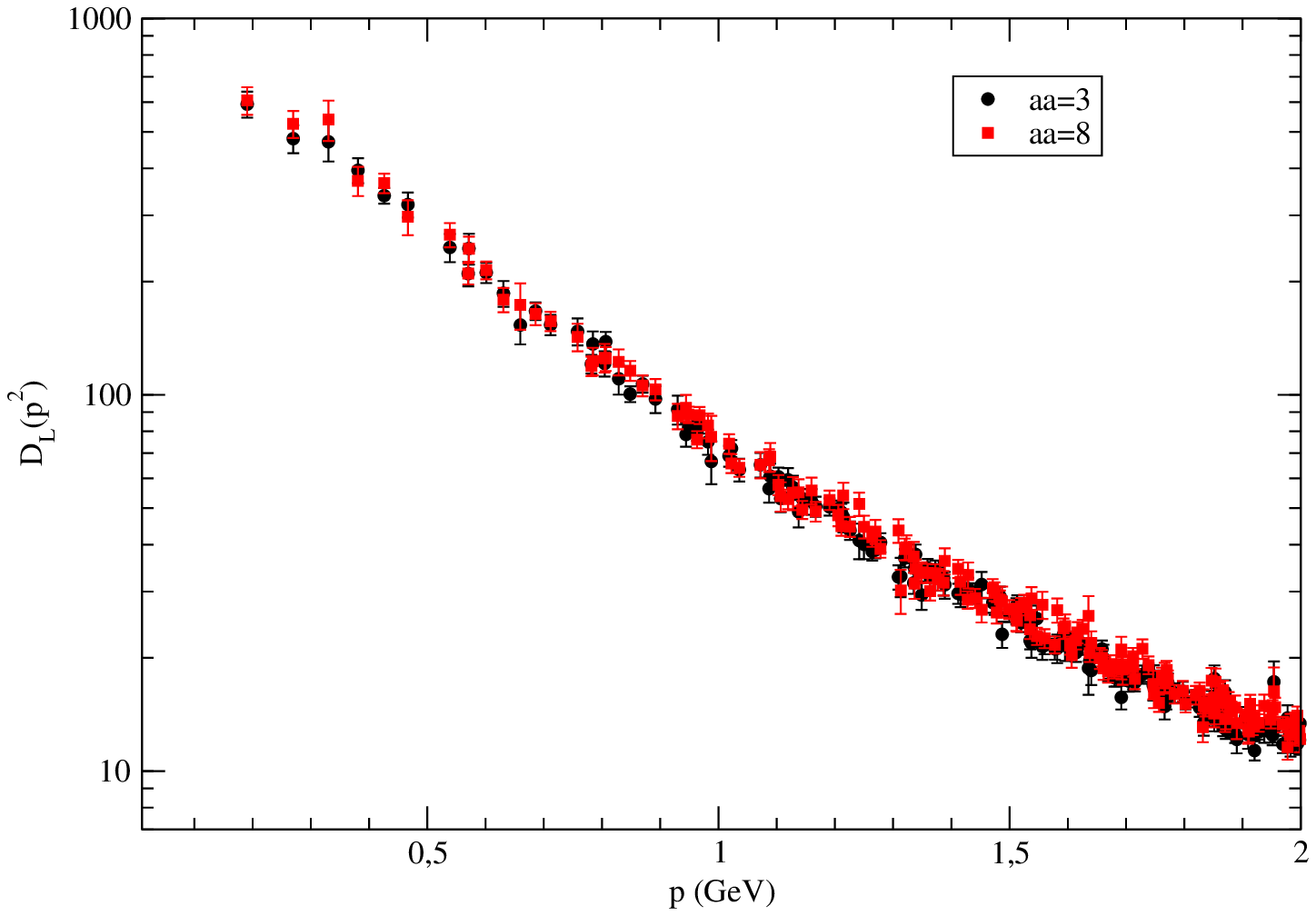} }
 \qquad
\subfigure{}{\includegraphics[angle=-90,width=0.42\textwidth]{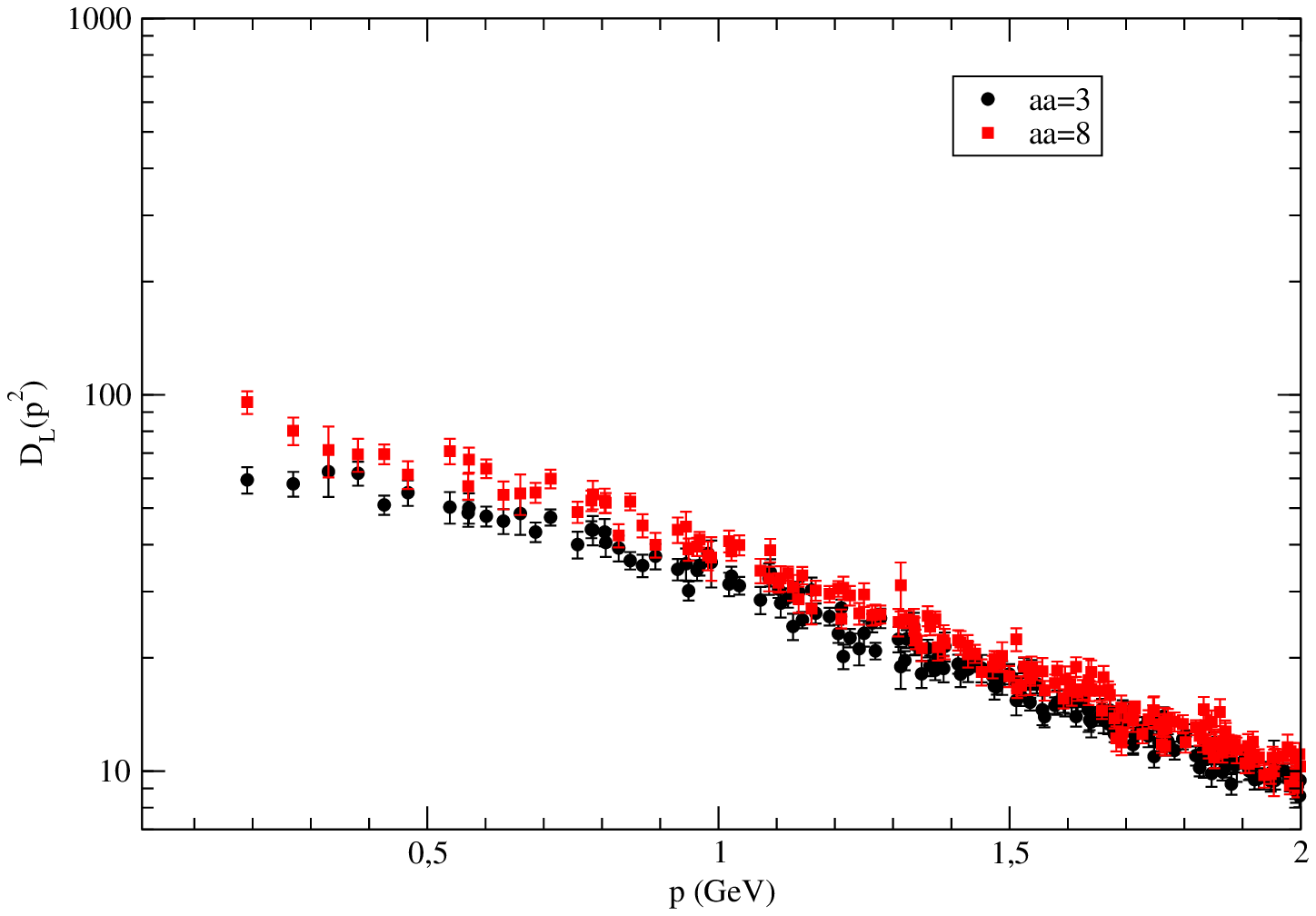} }

\caption{Bare gluon propagator for color indices 3 (black) and 8 (red), below $T_c$ (left) and above $T_c$ (right), for the various center sectors.}

\label{propagator}

\end{figure}

As shown in \cite{egmond1}, in the symmetric phase we expect that the propagators defined for color indices 3 and 8 coincide: $D^{33}=D^{88}$. This is exactly what we see on the left-hand  side plots of Fig. \ref{propagator}. For the broken phase, the propagators decouple, as seen in the right-hand plots.

\section{Conclusions}

Here we have briefly described the first steps of an ambitious research program towards a systematic lattice study of Green's functions in the center-symmetric Landau gauge. We have shown preliminary  lattice results in the center-symmetric Landau gauge for the link average and the gluon propagator. The numerical data supports the use of the link average and the difference of the color components of the gluon propagator as order parameters for the deconfinement phase transition. A detailed description of  the theoretical framework and updated results for the link average can be found in \cite{linkpaper}, and another paper is in preparation about the gluon propagator. Next steps will certainly include the ghost propagator, the $SU(2)$ case, and higher-order correlation functions.

\section*{Acknowledgements}

 This work was supported by FCT - Fundaç\~ao para a Ciência e a Tecnologia, I.P., under Projects Nos. UIDB/04564/2020 \cite{1}, UIDP/04564/2020 \cite{2} and CERN/FIS-PAR/0023/2021 \cite{3}. P.~J.~S. acknowledges financial support from FCT contract CEECIND/00488/2017 \cite{4}. P.~J.~S. would like to thank the financial support of CNRS and the hospitality of CPHT, École Polytechnique, where part of this work was performed. Computer simulations have been performed with the help of Chroma \cite{Edwards} and PFFT \cite{Pippig} libraries. The authors acknowledge the Laboratory for Advanced Computing at the University of Coimbra (\url{http://www.uc.pt/lca}) and the  Minho Advanced Computing Center (\url{http://macc.fccn.pt}) for providing access to the HPC resources. Access to Navigator was partly supported by the FCT Advanced Computing Projects 2021.09759.CPCA \cite{5}, 2022.15892.CPCA.A2 and 2023.10947.CPCA.A2. Access to Deucalion was supported by the FCT Advanced Computing Project 2024.11063.CPCA.A3.

\end{document}